\documentclass[doublecol,linenumbers]{epl2} % for 2 columns style with line numbers
\usepackage{mathptmx,amssymb}
\usepackage{amsmath}
\title{Experimental observation of 1/f noise in quasi-bidimensionnal turbulent flows}
\shorttitle{Title} %Insert here a short version of the title if it exceeds 70 characters

\author{J. Herault\inst{1} \and F. P\'etr\'elis\inst{1} \and S. Fauve\inst{1}}
\shortauthor{J. Herault \etal}

\institute{                    
  \inst{1} Laboratoire de Physique Statistique, Ecole Normale Sup\'erieure, CNRS, Universit\'e P. et M. Curie, Universit\'e Paris Diderot, Paris, France\\
}

\abstract{
We report the  experimental observation of $1/f^{\alpha}$ noise in quasi-bidimensionnal turbulence of an electromagnetically forced flow. The large scale velocity $U_L$ exhibits this power-law spectrum with $\alpha \simeq 0.7$ over a range of frequencies smaller than both the characteristic turn-over frequency and the damping rate of the flow. By studying  the statistical properties of sojourn time in each polarity of $U_L$, we demonstrate that the  $1/f^{\alpha}$ noise is generated by a renewal process, defined by  a two-state model given by the polarities of the large scale circulation. The statistical properties of this renewal process are shown to control the value of the exponent $\alpha$.}

\begin{document}

\maketitle

\section{Introduction}

Fluctuations which have spectral densities varying approximately as $1/f$ (or more generally as $1/f^{\alpha}$ with $0 < \alpha < 2$) over a large range of frequencies, or $1/f$ noise, have been studied since a long time in physics, first in the context of low frequency voltage fluctuations in electrical conductors~\cite{Dutta}. An early motivation has been the divergence problem related to a spectrum with a $1/f$ power law without any observed low frequency cut-off. Other questions concerned the non stationary or non Gaussian character of $1/f$ noise. The $1/f$ behavior has been first related to the existence of a broad band distribution of relaxation times in the system~\cite{Bernamont,vanderziel}. Other stochastic models include fractal Brownian motion~\cite{Mandelbrot1968} or power-law shot noise~\cite{Lowen1990}. Dynamical system theory provided another type of approach relying on deterministic low dimensional systems displaying a transition to chaos via intermittency~\cite{Manneville,Geisel}. These studies were useful to explain both the wide (power-law) distribution of relaxation times as well as the related $1/f^{\alpha}$ spectrum. This correspondence has been emphasized using purely stochastic models that involve random bursts or random switching between two states. It has been shown that if the interevent-time probability distribution function (PDF) decays as a power law, $P(\tau) \propto \tau^{-\beta}$, a $1/f^{\alpha}$ spectrum is obtained with $\alpha$ related to $\beta$~\cite{Lowen1993,Niemann}. 
Most of the early experimental observations of $1/f^{\alpha}$ noise do not display such discrete events. However, switching events have been observed in small electronic systems (submicrometer MOSFETs)~\cite{Ralls} and more recently in blinking quantum dots~\cite{Kuno}.

\begin{figure}[htb!]
\includegraphics[width=90mm,height=45mm]{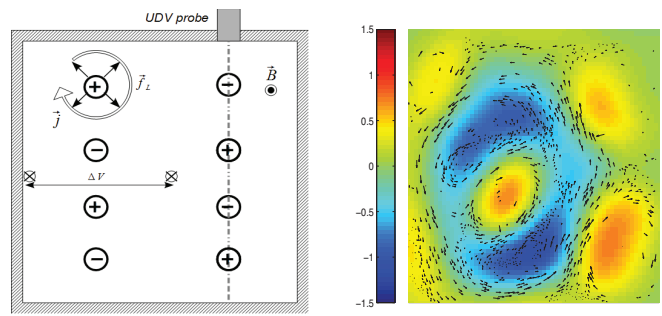}
\caption{\label{fig1} (Color online) Left : experimental set-up. The current injected through the electrodes and the vertically applied magnetic field induce an azimuthal Lorentz force close to each electrode. A pair of probes measure a potential difference $\Delta V$, proportional to the flow rate between the probes. The velocity component along the dashed line is measured using ultrasonic Doppler velocimetry. Right: Particle tracking of oxide tracers are used to compute the large scale vorticity levels in $s^{-1}$. Arrows represent the velocity field (Rh=23.5).}
\end{figure}

We present here the observation of a similar behavior on a macroscopic system, two-dimensional (2D) turbulence, where we show that switching events between the two polarities of the large scale circulation (LSC) account for the observed $1/f^{\alpha}$ noise. $1/f^{\alpha}$ spectra have been reported in various turbulent flows: in wall turbulence, they are observed in an intermediate frequency range and are ascribed to the $1/k$ spatial spectrum related to hairpin vortices~\cite{Perry1986}. They have been observed  for all frequencies below the integral scale in von Karman swirling flows, both for the pressure~\cite{Abry} and the velocity~\cite{Ravelet}. $1/f^{\alpha}$ spectra have been also observed in these flows for the fluctuations of the magnetic field, either when an external field is applied to a liquid metal~\cite{Bourgoin} or when the magnetic field is generated through the dynamo process~\cite{Monchaux}. Similar results have been found in numerical simulations of hydrodynamic or magnetohydrodynamic turbulence~\cite{Mininni}.

The experiment under study consists in  a quasi-bidimensional flow of a thin layer of liquid metal driven by a spatially periodic electromagnetic force. It has been predicted~\cite{Kraichnan} and experimentally checked~\cite{Sommeria,Paret} that in 2D turbulence, an inverse cascade of energy can drive a LSC superimposed on turbulent fluctuations. Whereas many studies have focused on the relation between coherent structures and spatial spectra, less attention has been paid to the frequency spectrum of 2D turbulent flows and to the possible signature of the dynamics of coherent structures. We study the temporal and spectral properties of the LSC and report the first experimental study of $1/f^{\alpha}$ noise in 2D turbulence. A striking feature of these fluctuations is their frequency range, which is well below the LSC turnover frequency and damping rate and extends to the lowest measured frequency without any low frequency cut-off. We explain how this $1/f^{\alpha}$ spectrum results from the dynamics of the LSC and we show that $\alpha$ is  related to the power-law exponent $\beta$ of the PDF of the waiting time between two successive changes of sign of the LSC.

\begin{figure}[htb!]
\begin{center}
\includegraphics[width=70mm,height=50mm]{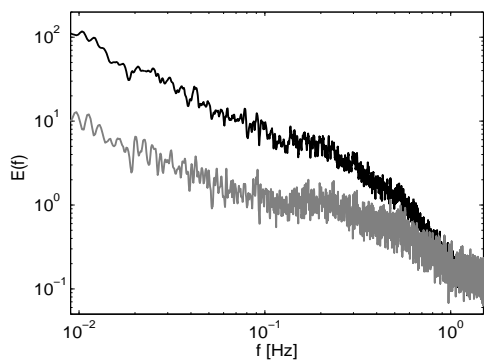}
\caption{Frequency power spectra $E(f)$ of one component of the velocity field for $Rh=26$: local velocity (grey), velocity averaged on the length of the cell (black).}
\label{dopplerspectra}
\end{center}
\end{figure} 

\section{Experimental set-up and techniques}

A thin layer of  liquid metal (Galinstan) of thickness $h=2\,$cm, is contained in a square cell of length $L=\, 12 cm$ submitted to a uniform vertical magnetic field up to $B_0 \simeq 0.1$T. A DC current $I$ (0-200A) is injected at  the bottom of the cell through an  array of 8 electrodes of diameter $d=8\, \hbox{mm}$ flush to the bottom of  the fluid layer (see Fig.1). In the vicinity of each electrode, the current density $\textbf j$ is radial so that the associated   Lorentz force $\textbf f_L=\textbf j \times \textbf B_0$  creates a  local  torque.  For low injected current, this forcing drives a laminar flow made of an  array of 8 counter-rotating vortices.  
Great care has been paid to inject the same current through  each electrode, in order to avoid the injection of net angular momentum in the flow. To prevent the oxidation of the upper surface, a thin layer of hydrochloric acid is placed on top of the liquid metal. 
Balancing  the Lorentz force and the inertia gives the typical velocity of the forced vortices, $U_c= \sqrt{ \vert \textbf f_L \vert L}$. Its order of magnitude is $10^{-1} m/s$ for $25 < I < 100$ A.  The bidimensionality of the flow is achieved  by a low magnetic Reynolds number, typically $Rm=\sigma \mu U_c L \sim 10^{-2}$, a relatively high interaction parameter $N=\sigma B_0^2 L / (\rho U_c) \sim 10$  and a high Hartmann Number $\hbox{Ha}= h B_0  [\sigma/(\rho \nu)]^{1/2} \sim 10^2$, where $\mu_0$,   $\sigma$ , $\nu$ are the magnetic permeability, electrical conductivity and kinematic viscosity respectively. 
In this parameter range, the velocity does not depend on the vertical coordinate except in a thin Hartman layer of size $\delta_h=h/\hbox{Ha}$ along the bottom boundary. This provides an additional dissipation to the 2D depth-averaged velocity field $\textbf v (x,y,t)$ that takes the form of  a  linear friction term in the 2D Navier-Stokes equation, namely $-\textbf v/\tau_H$ \cite{SommeriaM}, with a time scale $\tau_H=h \delta_H/\nu$ of the order of $10\,s$. We have checked that  $\tau_H$  is in good agreement with the experimentally measured damping rate of the large scale flow~\cite{herault}, confirming the bidimensionality of the flow as explained in \cite{SommeriaM,Shats}.

%At  the bottom of the cell, the competition between magnetic tension and viscous strain gives rise to an Hartman boundary layer of depth $\delta_h=h/\hbox{Ha}$. This Hartmann boundary layer provides an additional dissipation to the 2D depth-averaged velocity field $\textbf v (x,y,t)$. This dissipation  takes the form of  a  linear friction term in the 2D Navier-Stokes equation, namely $-\textbf v/\tau_H$ \cite{SommeriaM}, with a time scale $\tau_H=h \delta_H/\nu$ of the order of $10\,s$. The theoretical  value of $\tau_H$  is in good agreement with the experimentally measured damping rate of the large scale flow, confirming the bidimensionality of the flow \cite{Shats}. 

For the 2D flow,   we define two non-dimensional parameters from the two  sources of dissipation, viscosity and friction: the usual Reynolds number $Re=U_c L/\nu$ and $Rh=U_c \tau_H/L$ which is the ratio of inertia to linear friction. The ratio $Re/Rh$, independent of the injected current, is equal to  $\hbox{Ha} (L/h)^2 \sim 10^4$. By changing $I$ we vary $Rh$ between $1$ and $50$ so that  we reach relatively large Reynolds numbers.  Since viscous dissipation becomes efficient at scales smaller than $l = L \sqrt{Rh /R e} \sim 10^{-3} m$,  dissipation at large scale is mainly due to friction. It follows from these order of magnitude estimates that $Rh$ is the pertinent control parameter for the dynamics of the large scales, which is well verified experimentally~\cite{herault}. 

Velocity measurements are performed using three different methods~\cite{herault}: particle tracking (oxides on the Galinstan surface) shows the large scale velocity and the corresponding vorticity levels as displayed in Fig.1. An ultrasound transducer emits $4$ MHz wave trains along the dashed line in Fig. 1. They are reflected back by oxide particles in the flow and analyzed using a DOP3010 velocimeter (Signal Processing). The longitudinal velocity component is thus measured throughout the cell. The power spectra of the local velocity in the bulk of the cell and of the averaged velocity along the dashed line in Fig. 1, are shown in Fig. 2. They both display a power law behavior close to $1/f$ on a decade $0.01 < f < 0.1$ Hz. However, this measurement method is not well suited in the low frequency limit because measurements of long duration ($> 5$ hours) are difficult due to the large memory required to save and to process the data.
It has been known since a long time that a large scale velocity component can be directly determined by measurement the potential difference between a pair of electrodes in an external magnetic field~\cite{cramer2006}. As sketched  in Fig. 1, one of the electrodes is located in the middle of the cell and the other one close to the lateral wall, $5$ mm away from it. The potential difference between the electrodes $\Delta V$ is  $\Delta V \simeq \phi_L B_0$, with $\phi_L$  the flow rate between the center and the wall of the cell. In the following we use the spatially averaged velocity $U_L$, defined by $U_L=2 \phi_L/L$,   which is thus the large scale velocity coarse-grained on size $L/2$. 
% The aggregation of same-sign vortices generates a flow at the scale of the cell which corresponds to the LSC.  

\section{The different flow regimes and the low frequency spectrum}

Increasing $Rh$ from small values, coherent structures, with scales larger than the one of the forcing, are generated due to the non-linear energy transfers from small to large scales. 
For $Rh>5$, the flow is turbulent and several spatial scales contain energy as  displayed in Fig 1 (right). The dynamics of the LSC is chaotic. Its probability density function (PDF) is gaussian for intermediate values of $Rh$ but becomes bimodal for $Rh > 12$. The LSC then reverses between two values  $\pm U_L$ of maximum probability.   When $Rh$ is larger ($Rh \sim 30-40$), reversals of the LSC are less frequent and more visible on the direct recording of the velocity. They are no longer observed for $Rh >50$, for which the LSC has a constant sign. 
%This transition has been previously observed in a similar set-up but containing $36$ vortices \cite{Sommeria}.
All these flow regimes have been observed by numerical simulation of the 2D Navier-Stokes equation with damping~\cite{mishra2015}.  
The amplitude of the LSC given by the time series of $U_L$,  is displayed in  Fig. 3 for $Rh=17$. High frequency  turbulent fluctuations  are superimposed to  low frequency fluctuations. In particular we observe long events of constant sign (see for instance  $307 < t < 340$ s).  
 
\begin{figure}[htb!]
\begin{center}
\includegraphics[width=80mm,height=60mm]{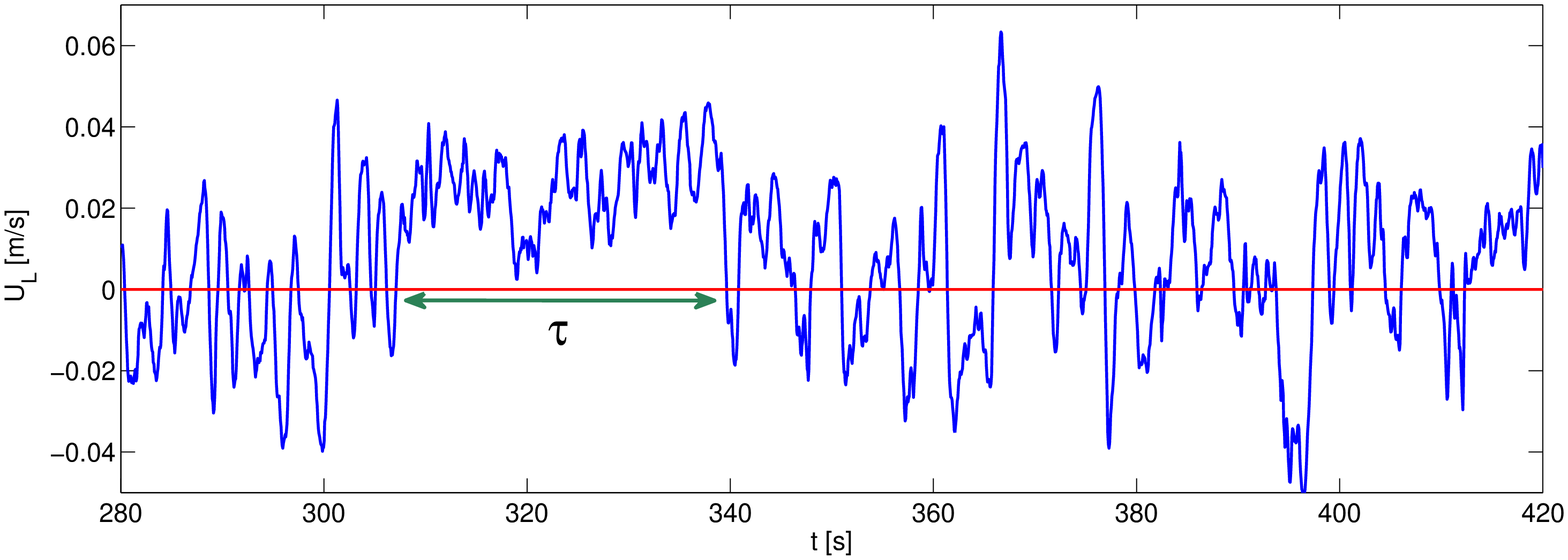}
\caption{ Time series of $U_L(t)$ for $Rh=17$. The velocity strongly fluctuates around zero, but events of constant polarity of duration  $\tau> L/ \langle \vert U_L \vert \rangle$ occur. }
\label{fig2}
 \end{center}
\end{figure}

To investigate the properties of the LSC, we calculate the temporal  power spectrum $E(f)$ of $U_L$ (see Fig. 4). Two distinct behaviors are observed (for $f$ larger or smaller than a crossover frequency $f_t$).  The spectrum is  steep for $f > f_t$. 
%Due to the small frequency range (smaller than a decade), we cannot determine whether $E(f)$  follows an  exponential or a power-law behavior  (close to  $f^{-4}$) for $f>f_t$. 
For $f<f_t$, the spectrum displays  a power-law behavior  $f^{-\alpha}$ with $\alpha=0.7$. 
%The compensated spectra $f^{0.7} E(f)$ are displayed in the inset. 
For $Rh<30$, the large scale flow thus exhibits  $1/f$ noise  over roughly two decades. For $Rh>30$, we observe a  departure from the $f^{-0.7}$ scaling. This phenomenon is related to the transition of the flow to the condensed regime where the statistical properties of the flow suddenly change~\cite{Kraichnan, Sommeria}.
% (standard deviation of $U_L$ and flow pattern). This transition has been characterized but deserves its own discussion.   
For small values of $Rh$ (between $5$ and $10$), the fluctuations of the LSC have a flat temporal power spectrum  at small frequency.  We thus restrict our study to $12< Rh <30$. We   report  the  values of the exponent $\alpha$ (blue circles)  as a function of $Rh$ in Fig. 5. The exponent is calculated on the interval $[f_c,f_t]$ where  $f_c/(2 \pi)$ is the inverse of the experiment duration which is  set by the thermal stability of the set-up, of the order of a few hours. 

\begin{figure}[htb!]
\begin{center}
\includegraphics[width=80mm,height=60mm]{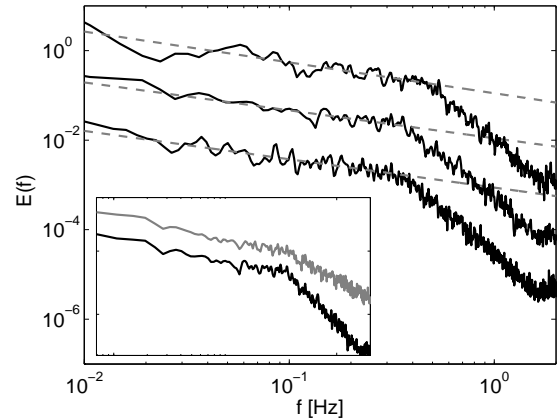}
\caption{Frequency power spectra $E(f)$ of $U_L(t)$ for  $Rh = 16,19, 24$, from the bottom to the top. Spectra have been shifted for clarity. The dashed lines are the best fits of the low frequency part of the spectra. Inset: frequency power spectra of $U_L $ (black), of the sign of $U_L $ (grey).
%compensated  power spectra  $f^{0.7} E(f)$.
 }
\label{fig3}
\end{center}
\end{figure}

For  frequencies larger than $f_c$, there is no sign of a frequency cut-off below which the spectrum would become flat. The maximum frequency $f_t$  is of order $0.4$ Hz and corresponds to several turnover times  of the LSC $ L/\langle \vert  U_L \vert \rangle$, with $\langle  \vert  U_L \vert \rangle \sim 2.10^{-2}m/s$,  a typical value of the time average of  $\vert U_L \vert$. 
The low frequency range corresponds to frequencies smaller than the slowest turnover time. Obviously there is no hope that a Taylor hypothesis could explain the spectrum  by an equivalence between frequency and wave number  $k$. In addition,  the $1/f$ range  extends to frequencies much smaller than the damping rate $1/\tau_h = 0.1$ Hz.  In other words,  the $1 /f$ spectrum is the frequency signature of large scale coherent structures, with life time larger than both their turnover time and the dissipation time $\tau_H$.

\begin{figure}[htb!]
\begin{center}
\includegraphics[width=80mm,height=50mm]{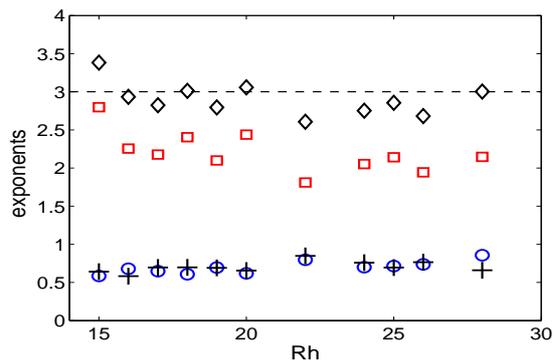}
\caption{(Color online) Exponents $\alpha$ ($\circ$) and  $\alpha_S$ ($+$) of the power law  of the  spectra $E(f)$ and $E_S(f)$. Exponent $\beta$ ($\square$) of the power law $P(\tau)$. $\alpha+\beta$ ($\diamond$) is nearly equal to $3$.  }
\label{fig5}
\end{center}
\end{figure} 

%\begin{figure}[htb!]
%\begin{center}
%\includegraphics[width=80mm,height=50mm]{spetrce_1surf_signe.eps}
%\caption{(Color online) Frequency power spectra of $U_L $ (middle, black), of the sign of $U_L $ (top, blue) and   of $\vert U_L \vert $ (bottom, red) for $Rh= 25$. The power spectrum of $\vert U_L \vert$ which is almost flat for low  frequencies while the power spectra of $U_L$ and of its sign are similar. }
%\label{fig4}
%\end{center}
%\end{figure}
%\vspace{-1cm}

\section{Relation between the low frequency spectrum and reversals of the large scale velocity}

As noticed in Fig. 3,  the LSC can maintain a  constant  direction for very long durations and  a natural question is the relation between these  events of constant polarity and the $1/f$ noise.  To what extent do these events control the low frequency part of the spectrum? To answer this question, we compute $E_S$, the power spectrum of the sign of $U_L$, thus keeping only the information of the direction of rotation of the LSC. $E_S$ is displayed in the inset of Fig. 4. It exhibits strong similarities with the spectrum of $U_L$ for $f<f_t$. These observations are confirmed by the calculation of the exponent $\alpha_S$ (see Fig. 5), defined by $E_S(f) \propto f^{-\alpha_S}$ for $f<f_t$. For all $Rh$, $\alpha$ and $\alpha_S$ are  almost equal, which implies that $E_S$ and $E$ contain the same spectral information for the low frequency range.

To further investigate  the information contained in the sign of $U_L$, we calculate the statistical properties of the time between sign changes. The PDF $P(\tau)$ of duration $\tau$ between two consecutive sign changes is shown in Fig. 6.  % for $Rh=  16 ,   19  ,  24$. 
$P(\tau)$ exhibits  a power law,  $P(\tau) \propto \tau^{-\beta}$ for  $\tau>4$ s, which corresponds to time scales larger than $f_t^{-1}$. The dashed line  is a power law with $\beta=2.25$. Such heavy-tailed distributions indicate that long durations have a high probability of occurrence. Then these events control the behavior of the autocorrelation function of the signal and thus are responsible for the form of the spectrum at low frequency.  %Signals that switch sign following an heavy tailed distribution for the interswitching time, have been considered for a large variety of applications, in particular in solid state physics (semiconductors, electronic noises...). In this context \cite{Lowen1993}, spectra have been calculated analytically for $P(\tau)$ similar to the ones that we measured experimentally. 
A simple argument can be stated as follows.  Let $u(t)$ be the renewal process defined by the sign of $U_L$. We note $T_i$ the interval between sign changes and consider the autocorrelation function 
 $C(\tau)=\langle u(t) u(t+\tau) \rangle $. We estimate $C(\tau)$ as $T_m^{-1} \int_0^{T_m} u(t) u(t+\tau) dt$, with $T_m$ tending to infinity.  This corresponds to averaging the signal with itself but shifted in time, so that only phases for which the signal is constant on duration $T_i$ larger than $\tau$ give a non zero contribution to $C(\tau)$. During $T_m$ there are $T_m P(T_i)  dT_i/ \langle T_i \rangle $ phases of duration $T_i$ which contribute 
$(T_i-\tau)$ to $C(\tau)$. We thus have $C(\tau) \propto \langle T_i \rangle^{-1} \int_0^{T_m} P(T_i) (T_i-\tau) dT_i$  which leads to $C(\tau)\propto \tau^{2-\beta}$. The associated spectrum follows from Wiener-Khinchin theorem, $E(f)\propto f^{\beta-3}$. We have thus obtained the relation  (see for instance \cite{Lowen1993})  $\alpha+\beta=3$.

To test this prediction,   we compute for different $Rh$ the exponent $\beta$ from $P(\tau)$   on the time interval $f_t^{-1} < \tau < f_c^{-1}$ corresponding to the frequency interval used to calculate the slope of the spectra. The exponents $\beta$ are displayed in Fig. 5 together with $\alpha+\beta$. %is displayed in Fig. 3 (black diamond). 
 We observe that  $\alpha+\beta$ is close to $3$. Our experimental measurements are thus in good agreement with the theoretical prediction. We  also stress that for $2 < \beta < 3$, the relation $\alpha=3-\beta$ still holds, even if the PDF of waiting time is asymmetric between  \textit{up} and \textit{down} states \cite{Lowen1993}. This implies that the relation is robust to possible imperfections of the experimental set-up, which could break the symmetry between the two  directions of rotation.

\begin{figure}[htb!]
\begin{center}
\includegraphics[width=80mm,height=50mm]{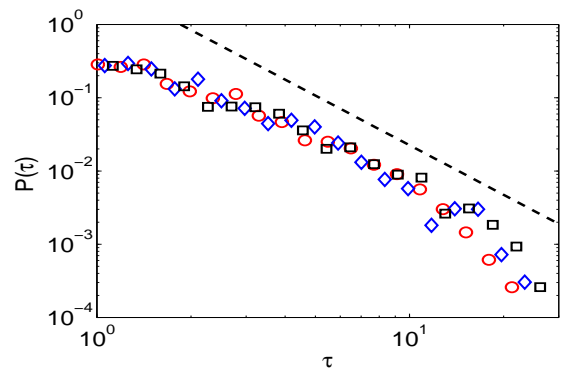}
\caption{(Color online) Probability density function $P(\tau)$ of duration $\tau$ between two consecutive sign changes  for $Rh=  16\, (\circ) ,   19 \, (\square)  ,  24 \, (\diamond)$. Dashed line: $\tau^{-\beta}$ with $\beta=2.25$}
\label{fig6}
\end{center}
\end{figure}  
 
%{\bf si on a la place, on peut peut etre rajouter des figures du champ de vitesse dans les phases de longue duree}

%{\bf on devrait probablement aussi brievement discuter les mesures doppler} 

\section{Conclusions}

We conclude  that the $1/f^{\alpha}$ spectrum is related to the power-law scaling of the PDF of the sojourn time in each polarity of the large scale flow. This observation raises two questions: first, are the $1/f^{\alpha}$ spectra observed in other turbulent flows also related to the statistical properties of coherent structures? Second, what is the origin of the power law distributions of lifetimes of large scale structures in 2D turbulence or other turbulent flows?

Motivated by our observations on 2D turbulence, we considered data for the pressure fluctuations in 3D turbulence~\cite{Abry}. 
We confirmed that the $1/f^{\alpha}$ spectrum of pressure is related to the power-law scaling of the waiting time between successive pressure drops due to intermittent vorticity filaments. We made a similar observation for the $1/f^{\alpha}$ spectrum of the magnetic field generated by a dynamo process~\cite{Bourgoin} that results from the statistics of bursts of magnetic field. We also believe that the $1/f$ spectrum of the velocity in the von Karman flow \cite{Ravelet} results from the power-law scaling of the switching dynamics of the shear layer. The analysis of all these data will be reported elsewhere~\cite{Heraultdsfd}.
The second question related to the origin of the long lifetimes of the LSC and the power-law scaling of the sojourn time in each polarity is still open. 
%This is related to the stability of the LSC and to the underlying mechanisms of the inverse cascade. Recent progresses have been made to understand the scale-to-scale fluxes of energy \cite{Xia,Kelley}, and could be a starting  point to investigate the stability of the LSC. However, the non-local interactions between the forcing and the  large scale structures also play an important role in the stability of the LSC  \cite{Tsang, Gallet}. 
We have observed that the switching dynamics between LSC of opposite polarities is rather complex and involve several intermediate states with different vorticity distributions. Thus, we can assume that the whole switching process requires the successful completion of several independent transitions, a mechanism that is known to generate power law distributions~\cite{Montroll}.


\begin{thebibliography}{0}

\bibitem{Dutta}
  \Name{Dutta P. \and Horn P. M.}
  \REVIEW{Rev. Mod. Phys.}{53}{1981}{497}.
    
\bibitem{Bernamont}
  \Name{Bernamont J.}
  \REVIEW{Proc. Physical Soc.}{49}{1937}{138}.
  
\bibitem{vanderziel}
  \Name{van der Ziel A.}
  \REVIEW{Physica}{16}{1950}{359}.
  
\bibitem{Mandelbrot1968}
  \Name{Mandelbrot B. B. \and van Ness I. W.}
  \REVIEW{SIAM Rev.}{10}{1968}{422}.
  
\bibitem{Lowen1990}
  \Name{Lowen S. B. \and Teich M. C.}
  \REVIEW{IEEE Trans. Info. Theory}{36}{1990}{1302}.
  
\bibitem{Manneville}
  \Name{Manneville P.}
  \REVIEW{J. Physique}{41}{1980}{1235}.
  
\bibitem{Geisel}
  \Name{Geisel T., Zacherl A. \and Radons G.}
  \REVIEW{Phys. Rev. Lett.}{59}{1987}{2503}.  
  
\bibitem{Lowen1993}
  \Name{Lowen S. B. \and Teich M. C.}
  \REVIEW{Phys. Rev. E}{47}{1993}{992}.

\bibitem{Niemann}
  \Name{Niemann M., Kantz H. \and Barkai E.}
  \REVIEW{Phys. Rev. Lett.}{110}{2013}{140603}.

\bibitem{Ralls}
  \Name{Ralls K. S. et al}
  \REVIEW{Phys. Rev. Lett.}{52}{1984}{228}.

\bibitem{Kuno}
  \Name{Kuno M. et al}
  \REVIEW{J. Chem. Phys.}{112}{2000}{3117}.

\bibitem{Perry1986}
  \Name{Perry A. E., Henbest S. \and Chong M. S.}
  \REVIEW{J. Fluid. Mech.}{165}{1986}{163}.

\bibitem{Abry}
  \Name{Abry P. et al}
  \REVIEW{J. Physique II}{4}{1994}{725}.
  
\bibitem{Ravelet}
  \Name{Ravelet F., Chiffaudel A. \and Daviaud F.}
  \REVIEW{J. Fluid. Mech.}{601}{2008}{339}.  
  
\bibitem{Bourgoin}
  \Name{Bourgoin M. et al}
  \REVIEW{Phys. Fluids}{14}{2002}{3046}.

\bibitem{Monchaux}
  \Name{Monchaux R. et al}
  \REVIEW{Phys. Fluids}{21}{2009}{035108}.
  
\bibitem{Mininni}
  \Name{Mininni P. et al}
  \REVIEW{Phys. Rev. E}{89}{2014}{053005} and references therein.
  
\bibitem{Kraichnan}
  \Name{Kraichnan R. H.}
  \REVIEW{Phys. Fluids}{10}{1967}{1417}.
  
\bibitem{Sommeria}
  \Name{Sommeria J.}
  \REVIEW{J. Fluid. Mech.}{179}{1986}{139}.   

\bibitem{Paret}
  \Name{Paret J. \and Tabeling P.}
  \REVIEW{Phys. Fluids}{10}{1998}{3126}.
  
\bibitem{SommeriaM}
  \Name{Sommeria J. \and Moreau R.}
  \REVIEW{J. Fluid. Mech.}{118}{1982}{507}. 
  
\bibitem{herault}
  \Name{Herault J.}
  \Book{PhD thesis: Dynamique des structures coh\'erentes en turbulence magn\'etohydrodynamique}
  \Publ{Universit\'e Paris-Diderot}
  \Year{2013}.

\bibitem{Shats}
  \Name{Shats M., Byrne D. \and Xia H.}
  \REVIEW{Phys. Rev. Lett.}{105}{2010}{264501}.
  
\bibitem{cramer2006}
  \Name{Cramer A. et al}
  \REVIEW{Flow Meas. Instrum.}{17}{2006}{1}.

\bibitem{mishra2015}
  \Name{Mishra P. et al}
  \REVIEW{Phys. Rev. E}{91}{2015}{053005}.

\bibitem{Heraultdsfd}
  \Name{Herault J., P\'etr\'elis F. \and Fauve S.}
   \REVIEW{J. Stat. Phys.}{161}{2015}{16}  
  
%\bibitem{Tsang}
  %\Name{Tsang Y.-K. \and Young W. R.}
  %\REVIEW{Phys. Rev. E}{79}{2009}{045308}.

%\bibitem{Gallet}
  %\Name{Gallet B. \and Young B.}
  %\REVIEW{J. Fluid. Mech.}{715}{2013}{359}.   
  
\bibitem{Montroll}
  \Name{Montroll E. W. \and Shlesinger M. F}
  \REVIEW{Proc. Natl. Acad. Sci. USA}{79}{1982}{3380}.   

\end{thebibliography}
\end{document}